# Surface dipole of F4TCNQ films: Collective charge transfer and dipole-dipole repulsion in submonolayers

Zoltán G. Soos and Benjamin J. Topham

Department of Chemistry, Princeton University, Princeton NJ 08544

## Abstract

A charge transfer (CT) model is introduced for strong acceptors like A = F4TCNQ that are ionized on surfaces at low coverage $\theta$. Each A forms a CT dimer with a surface state S. Dipole-dipole repulsion grows as $\theta^{3/2}$ up to $\theta = 1$ in a full monolayer. Electron transfer $\rho(\theta)$ within dimers is found self-consistently and decreases with increasing coverage. The surface dipole and work function shift $\Delta\Phi$ are proportional to $\theta\rho(\theta)$. The CT model has implications for photoemission and accounts for $\Delta\Phi(d)$ profiles of F4TCNQ films of thickness d on Cu(1,1,1) or on hydrogenated diamond (1,0,0). The CT model also describes $\Delta\Phi(\theta)$ of organic donors on metals and is contrasted to previous treatments of dipoles on a surface.

## I. Introduction

F4TCNQ (tetrafluorotetracyanoquinodimethane) is a powerful π-electron acceptor (A), even stronger than TCNQ. TCNQ forms mixed stacks with π-donors (D) and an extensive series of salts with segregated stacks of $A^-$ or $(A_2)^-$ anion radicals and closed-shell cations between stacks.[1,2] As reviewed by Chen *et al.*,[3] recent interest in F4TCNQ films has been in the context of organic electronics[4-6] and promoting hole injection[7] by modulating the work function via surface dipoles. Initial adsorption of F4TCNQ on semiconductors or metals is as $A^-$ anion radicals as shown by two diagnostic electronic states, $E_1$ and $E_2$, separated by ~1.2 eV in photoemission.[8,9] Adsorption beyond a monolayer (ML) involves neutral molecules A with photoemission at higher binding energy $E_3$ and suppression of $E_1$ and $E_2$. Indeed, surface dipole[3] and other data[10] indicate that $A^-$ coverage saturates below a ML. There is broad agreement about the principal features of F4TCNQ films. There are different estimates of the saturation coverage of $A^-$, however, or how adsorption evolves from $A^-$ to A. Moreover, agreement about $A^-$ on surfaces does not extend to cations $D^+$ when the adsorbed molecule is a donor. Photoemission evidence is much less accessible because $D^+$ has higher binding energy than D. Nevertheless, initial adsorption of D = TTF (tetrathiafulvalene) on Au(111) is also as $D^+$ leading to ordered sub ML structures that were attributed to dipole-dipole repulsion.[11]

Metal-organic (m-O) interfaces have been extensively studied in connection with organic electronics applications, as reviewed by Ishii *et al.*[12] and Hwang *et al.*[13] A major advance has been to identify surface dipoles $\Delta(d)$ that are formed or almost formed at film thickness $d = d_{ML}$. By contrast, interfaces between metals and wide band inorganic semiconductors show band bending over a length scale that is an order of magnitude longer. There is a fundamental difference between metal-semiconductor and m-O interfaces. Possible physical origins of $\Delta$ have been discussed.[12,13] Density functional theory (DFT) and quantum chemical calculations have mainly addressed the magnitude of $\Delta \sim \Delta(d_{ML})$. The evolution of $\Delta(d)$ with $d \leq d_{ML}$ has not been modeled in general, to the best of our knowledge. In this paper, we model Δ(d) profiles of sub ML F4TCNQ films by a method that is applicable to any m-O interface with weak nonspecific bonding. Profiles yield information about the charge of adsorbed species and discriminate among possible contributions to surface dipoles.

The phenomenological model of Δ(d) profiles is presented in Section 2. Here we introduce the principal ideas. As sketched in Fig. 1, we suppose that each adsorbed A forms a charge transfer (CT) dimer with a surface state S of the metal. Following Mulliken's treatment of DA dimers, the wave function is a linear combination of a neutral state $|AS\rangle$ and an ionic state $|A^-S^+\rangle$ with dipole $\mu_0 < 0$ normal to the surface. The coefficient $\rho(\theta)^{1/2}$ of the ionic state gives the fraction $\rho(\theta)$ of transferred charge. The separation $r(\theta)$ between adsorbed molecules decreases with increasing $\theta = d/d_{ML} \leq 1$, and dipole-dipole repulsion reduces $\rho(\theta)$. Collective CT refers to the self-consistent treatment of dimers on an infinite surface. We follow previous CT studies of organic crystals[14,15] or molecular aggregates.[16,17] The energies of the two states are $E(AS)$ and $E(A^-S^+)$ when the CT integral is $t = 0$. Although $t$ is small by hypothesis, strong mixing is possible when $E(AS) \sim E(A^-S^+)$ are nearly degenerate, as required for neutral-ionic transitions or crossovers.[15,17]

In addition to surface dipoles, m-O interfaces typically have an induced density of interfacial states (IDIS) as shown by deviations from the Schottky-Mott limit.[12,13] Induced states are consistent with CT. Weak nonspecific bonding is thought[13] to describe m-O interfaces in contrast to covalent bonds in MLs of alkanethiols on Au or at chemically modified surfaces. Chemisorption also involves strong interactions with $t > 1$ eV in the Newns model.[18] While there is no sharp boundary between weak and strong interactions, CT models assume small $t$ and emphasize electron correlation. DFT or direct calculations are more detailed but less correlated descriptions. Quite generally, dipoles have been associated with atoms or molecules adsorbed on metals, semiconductors or insulators, and qualitative aspects of interfaces are well known. Dipole models have also been criticized and are best for weak interactions.

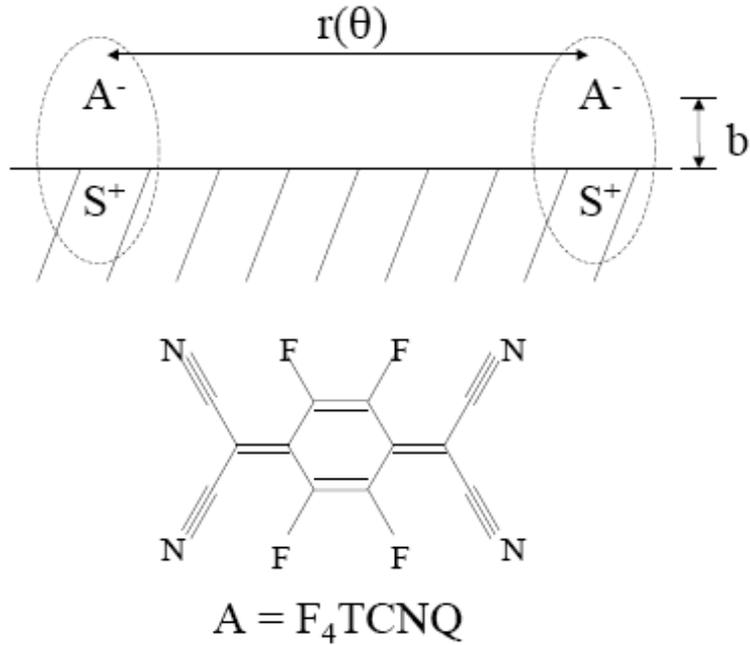

Fig. 1. Schematic representation of adsorbed F4TCNQ molecules on a surface separated by r(θ) at coverage θ ≤ 1. Each A forms a charge transfer (CT) dimer with a surface state S.

The Topping model also describes nonspecific bonding that can be applied to Δ(d) profiles.[19,20] The model assumes a fixed dipole $p_0$ and polarizability α for every adatom. The dipole field F(θ) at coverage θ is found self-consistently for induced dipoles αF(θ) that reduce the net dipole to p(θ). An example of a quantitative fit[19] is the Δ(d) profile of Cs on Si(1,1,1). Fixed $p_0$ and depolarization is quite different from variable dipoles and dipole-dipole repulsion in the CT model. But the sharp mathematical difference is less sharp physically since both polarization and CT involve charge redistribution. The models describe work function shifts that cannot uniquely be partitioned into CT and polarization. Since different profiles are predicted, however, observed shifts may point to either model. CT is the natural starting point for F4TCNQ films in view of evidence for ions at θ ~ 0 and molecules at θ ~ 1. We are applying[21] the CT model to published Δ(d) profiles with CT from molecule to metal and find equally good agreement for $D^+$ ions at $\theta \sim 0$.

The CT model of interacting dimers is presented in Section 2. Its parameters are $\rho(0)$ and $\rho(1)$, the initial and final degree of CT. Alternatively, the parameters are $E(A^-S^+) - E(AS)$ and dipole-dipole repulsion V, both in units of $t = 1$. Two scale factors enter the work function shift $\Delta\Phi(d)$, the ML thickness $d_{ML}$ and the shift $\Phi_0$ of an ionic ($\rho = 1$) ML. Section 3 has applications to F4TCNQ films. Calculated $\Delta\Phi(d)$ profiles are compared to work function shifts of F4TCNQ films on hydrogenated diamond (1,0,0) and Cu(1,1,1) surfaces. The Discussion contrasts the Topping and CT models and comments briefly on direct approaches to m-O interfaces. We mention extensions of the CT model and emphasize its phenomenological nature.

**2. Collective CT model of submonolayers**

An isolated A on the surface is modeled as a CT dimer with a localized surface state S whose precise nature is left open. The neutral state $|AS\rangle$ has energy E(AS) and dipole $\mu_N$ that we neglect for simplicity. The ionic singlet state $|A^-S^+\rangle$ has energy $E(A^-S^+)$ and large dipole $\mu_0 = -eb$ normal to the surface with $b \sim 3.5$ Å for a planar conjugated molecule in the prone position. The energy difference is $2\Gamma(0) = E(A^-S^+) - E(AS)$ in the limit $t = 0$. The two states are coupled by a small CT integral $t = \langle A^-S^+|H|AS\rangle/\sqrt{2}$ that does not perturb the electronic structure of A and $A^-$ significantly. The ground state (gs) of this 2x2 configuration interaction (CI) problem is[14]

$$|\psi(0)\rangle = \sqrt{1-\rho(0)}|AS\rangle + \sqrt{\rho(0)}|A^-S^+\rangle. \tag{1}$$

The ratio $t\sqrt{2}/\Gamma(0) = \tan 2\gamma$ and $\rho(0) = \sin^2\gamma$ determines the relative neutral and ionic contributions. Without loss of generality we set $t = 1$ as the unit of energy and obtain

$$\varepsilon(0) = -2\sqrt{2\rho(0)(1-\rho(0))} + 2\rho(0)\Gamma(0) \tag{2}$$

for an isolated dimer. The parameter $\Gamma(0)$ fixes $\rho(0)$, with $\Gamma(0) < 0$ for $\rho(0) > 1/2$.

F4TCNQ covers $A_0 \sim 100$ Å$^2$ in Fig. 1, so that a ML has $\sim 10^{14}$ acceptors/cm$^2$. We consider decreasing electron transfer $\rho(\theta) < 1$ with increasing $\theta$ due to dipole-dipole repulsion. The nearest-neighbor separation scales as $r(\theta) = r_0 \theta^{-1/2}$ for an invariant surface structure. The model Hamiltonian for $0 < \theta \leq 1$ is

$$H(\theta) = \sum_p h_p + \tfrac{1}{2} \sum_{pp'}{}' V_{pp'} \rho_p(\theta) \rho_{p'}(\theta) \tag{3}$$

where $h_p$ describes an isolated dimer and $V_{pp'} = \mu_0^2/r_{pp'}^3$ is the dipole-dipole interaction. We introduce a BCS-type trial function for $N_d = N\theta$ dimers at coverage $\theta$

$$|\psi(\theta)\rangle = \prod_{p=1}^{N_d} \left(\sqrt{1-\rho(\theta)} + \sqrt{\rho(\theta)}\, b_p^+\right)|N\rangle. \tag{4}$$

$|N\rangle$ is a product of $N\theta$ neutral dimers whose interactions with the surface or with each other are neglected. The operator $b_p^+$ acts on the pth dimer to generate the singlet $|A_p^- S_p^+\rangle$. The trial function contains all possible distributions of $|AS\rangle$ and $|A^-S^+\rangle$ states on the surface, and it is strongly peaked at $\rho(\theta)$. CI mixing $\rho(\theta)$ is the same for all dimers by construction. As in BCS, there is perfect correlation of two states ($k\alpha$, $-k\beta$ there, electron-hole excitation of dimers here) and a mean-field approximation otherwise.

The gs expectation value of $|\psi(\theta)\rangle$ is

$$E(\theta) = N_d \varepsilon(\theta) + \rho(\theta)^2 \sum_{pp'}{}' V_{pp'}/2. \tag{5}$$

The first term is Eq. 2 with $\rho(\theta)$ instead of $\rho(0)$. The second term contains sums over $1/r_{pp'}^3$ that scale as $\theta^{3/2}$

$$\sum_{p,p'} V_{pp'} = 2kN_d\mu_0^2\theta^{3/2}/r_0^3 \equiv 2V_0\theta^{3/2}N_d. \tag{6}$$

Topping[20] evaluated the sum accurately and obtained $2k \sim 9$ per unit area of either a hexagonal or square planar lattice. $2V_0$ is the repulsion per dimer in a ML with $\rho(1) = 1$. Minimization of $E(\theta)$ with respect to $\rho(\theta)$ gives the self-consistency condition

$$\frac{1-2\rho(\theta)}{\sqrt{2\rho(\theta)(1-\rho(\theta))}} = \Gamma(0) + V\theta^{3/2}\rho(\theta) \equiv \Gamma(\theta) \tag{7}$$

with $V = V_0/t$. A dimer with $\rho(0) > 1/2$ and $\Gamma(0) < 0$ at $\theta = 0$ becomes less ionic with increasing $\theta$. The states are degenerate at $\theta_1$, where $\rho(\theta_1) = 1/2$, $\Gamma(\theta_1) = 0$ and

$$V\theta_1^{3/2} = -2\Gamma(0) \tag{8}$$

The relation between V and $\Gamma(0)$ is independent of t. Alternatively, the model's two parameters may be taken as $\rho(0)$ and $\rho(1)$, the initial and final CT in Eq. 7.

The surface dipole is $\mu_0\theta\rho(0)$, linear in coverage for $V = 0$. Repulsion $V > 0$ gives nonlinear profiles and superlinear $\theta^{3/2}$ may even produce a $\theta\rho(\theta)$ maximum. The derivative

$$\mu'(\theta) = \mu_0\rho(\theta)\left(\frac{(2+\Gamma(\theta)^2)^{3/2} - V\theta^{3/2}/2}{(2+\Gamma(\theta)^2)^{3/2} + V\theta^{3/2}}\right) \tag{9}$$

vanishes at the $\theta\rho(\theta)$ maximum. Since $\Gamma(\theta_1) = 0$, we have $\Gamma(0) = -2\sqrt{2}$ and $\rho(0) = 0.9476$ when $\mu'(\theta_1) = 0$. The maximum shifts to $\theta < \theta_1$ for $\Gamma(0) < -2\sqrt{2}$ and to $\theta > \theta_1$ for $\Gamma(0) > -2\sqrt{2}$. There is no maximum when the numerator of Eq. 9 is positive at $\theta = 1$.

The surface dipole is $b\sigma_-(\theta)$, with charge density $\sigma_-(\theta) = -e\theta\rho(\theta)/A_0$. The change of electrostatic potential across the acceptor layer is $-4\pi eb\sigma_-(\theta)$. The work function shift is

$$\Delta\Phi(\theta) = -4\pi eb\sigma_- \equiv \Phi_0\theta\rho(\theta). \tag{10}$$

The prefactor $\Phi_0 = -4\pi e\mu_0/A_0$ is the shift a full ML of $A^-$ ions with $\mu_0 = -eb$. Comparison with experiment requires $\Phi_0$ and $d_{ML}$. First we consider the CT model itself. Figure 2 shows $\theta\rho(\theta) = \mu(\theta)/\mu_0$ for the indicated $\theta_1$ and $\rho(0) \sim 1$. Curves with $\rho(0) > 1/2$ cross at $\theta_1$. Increasing $\theta_1$ amounts to decreasing dipole-dipole repulsion V in Eq. 8. The surface dipole is $\sim\mu_0\theta_1/2$ over a broad range with a maximum at $\theta = \theta_1$ for $\rho(0) = 0.95$ as expected from Eq. 9 and at $\theta < \theta_1$ for $\rho(0) = 0.98$. Saturation at $\sim\theta_1/2$ is sudden because the repulsion $\theta^{3/2}$ in Eq. 6 is superlinear. The change becomes more collective with increasing $\rho(0)$.

Figure 2 also shows $\theta\rho(\theta)$ for largely neutral dimers with $\Gamma(0) \to -\Gamma(0)$, and hence $\rho(0) \to 1 - \rho(0) = 10\%$, and unchanged repulsion V based on $\theta_1$. The surface dipole is small and nearly linear, with little indication of cooperative behavior at $\rho(0) = 10\%$ and even less for $\rho(0) < 10\%$. We have neglected the dipole $\mu_N$ of $|AS\rangle$ compared to $\mu_0$. The model is readily generalized to include $\mu_N$ as another parameter. It is then convenient to redefined $\mu_0 \to \mu_0 - \mu_N$ and the self-consistent analysis holds with minor changes. Our reasons for setting $\mu_N = 0$ are that there are other contributions to $\Delta(d)$ and that CT addresses CI while $\mu_N$ reflects local bonding.

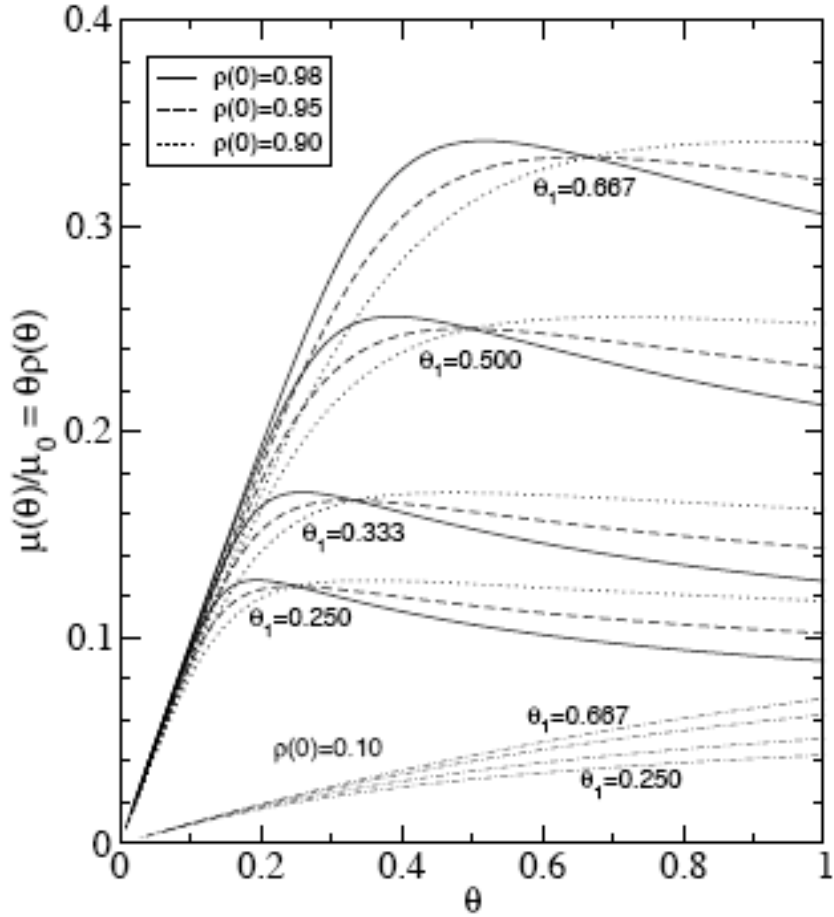

Fig. 2. Surface dipole $\mu(\theta)/\mu_0 = \theta\rho(\theta)$, Eq. 10, at coverage $\theta \le 1$. The dipole $\mu_0\rho(0)$ of isolated dimer decreases to $\mu_0/2$ at $\theta_1$ due to repulsion V in Eq. 7; increasing V decreases $\theta_1$ in Eq. 8. Maxima of $\theta\rho(\theta)$ are given by Eq. 9. Small $\rho(0) = 0.10$ surface dipoles with the same $\theta_1$ are nearly linear in $\theta$.

## 3. F4TCNQ on metals or semiconductors

F4TCNQ films have been discussed in terms of ions $A^-$ on the surface and molecules A in thick films. The growth mode is not known, but a full ML or a thick film surely has more adsorbed molecules than the ~20% coverage of $A^-$ inferred from $\Delta\Phi$ shifts.[3,7,10,22,23] CT dimers are linear combinations of $A^-$ and A according to $|\psi(\theta)\rangle$ in Eq. 4, with $\rho(\theta)$ found self-consistently via Eq. 7 to yield the profile $\Delta\Phi(\theta)$ in Eq. 10. There are several ways to compare with experiment: (1) the magnitude of $\Delta\Phi(d)$ for films of thickness d; (2) $\Delta\Phi(d)$ profiles as a function of d with $d_{ML} \sim 4$ Å for F4TCNQ; (3) photoemission spectra or other probes that identify surface species; (4) the surface roughness of sub ML films, which is open at present. The generality of the CT model invites comparison with sub ML films of any m-O interface.

Figure 3 shows published $\Delta\Phi(d)$ profiles to d = 10 Å for F4TCNQ on Cu(1,1,1) from Fig. 2 of ref. 22 and for the surface dipole contribution on hydrogenated diamond (1,0,0) from Fig. 3.5b of ref. 3. To calculate profiles, we took $\Phi_0 = -4\pi e\mu_0/A_0 = 6.3$ eV in Eq. 10 with $A_0 = 100$ Å$^2$ and $\mu_0 = -eb$ with b = 3.5 Å. Such inputs pertain to the interface rather than the model. The calculated profiles have $\rho(0) = 0.50$ and $\rho(1) = 0.09$ in Eq. 7, a five fold decrease in the ML, or alternatively $\Gamma(0) = 0.0$ and V = 22.5. The shift $\Delta\Phi(d_{ML}) = \Phi_0\rho(1)$ gives an estimated $\rho(1)$. The initial slope of $\Phi_0\rho(0)/d_{ML}$ is an estimate of $\rho(0)$. Shifts beyond d = $d_{ML}$ are small (~0.10 eV), comparable to experimental uncertainties, and not modeled. The fit demonstrates the viability of the model. Moreover, F4TCNQ/Ag(1,1,1) has[23] $\Delta\Phi \sim 0.65$ eV and the larger $\Delta\Phi = 1.3$ eV on a passivated Si surface,[9] comparable to diamond (1,0,0), was not resolved into surface dipole and band bending. Similar $\Delta \sim \Delta(1)$ for F4TCNQ on different surfaces points to nonspecific bonding.

Photoemission indicates TCNQ$^-$ or F4TCNQ$^-$ on surfaces. Grobman *et al*.[8] assigned TCNQ$^-$ photoemission at energy $E_1$ to ionization of the anion's singly occupied HOMO, the neutral molecule's LUMO. The next emission at $E_2$ was assigned to a "relaxed HOMO" level, perturbed by the partially filled LUMO. This assignment is regrettably still current,[3,22] long after Masuda *et al*.[10] pointed out the proper interpretation of $E_2$. Photoemission of an anion generates the electronic spectrum of the neutral molecule. The final state for $E_2$ emission is the lowest triplet of TCNQ or F4TCNQ. The same MOs give a singlet excited state at higher binding energy. With resolution limited to ±0.1 eV at best, the reported values of $E_2 - E_1$ for F4TCNQ are 1.3 eV on graphite,[10] 1.4 eV on passivated[9] Si, 1.0 eV on polycrystalline[7] Au, 1.2 eV on[2] Ag(1,1,1), 1.0 eV on hydrogenated diamond[3] (1,0,0), and 0.9 eV on Na-doped[24] Al$_2$O$_3$. Density functional theory (B3LYP) with the reasonably large 6-311++G** basis in the GAUSSIAN 03 program[25] returns 1.09 eV for the vertical triplet state of F4TCNQ.

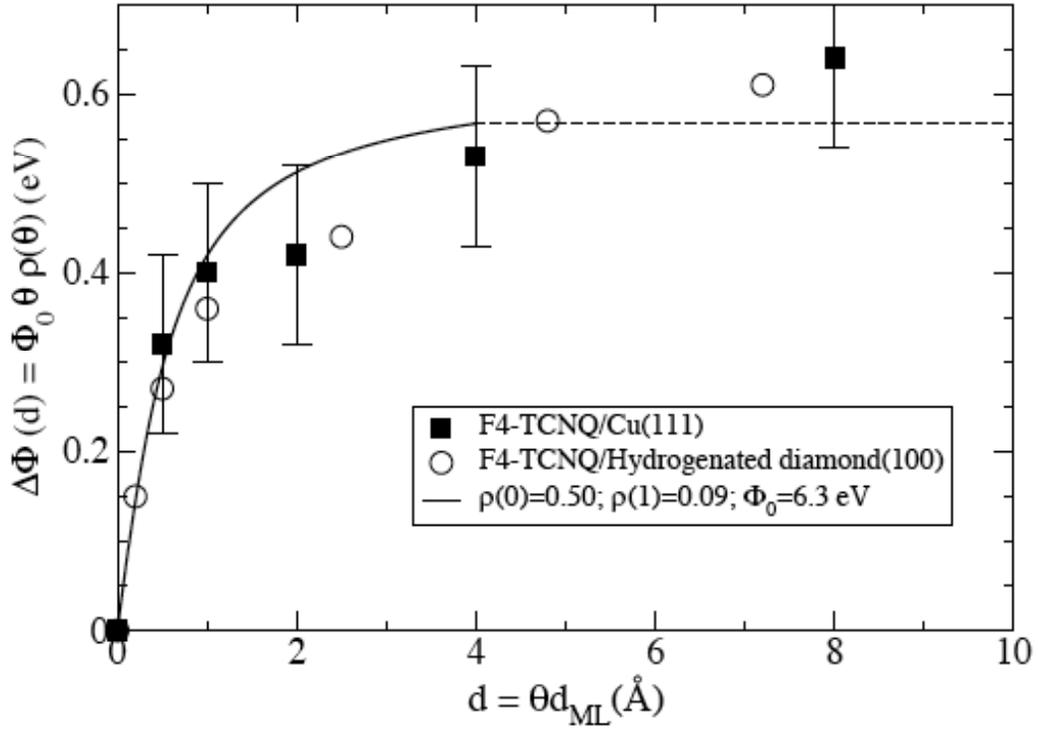

Fig. 3. Work function shift $\Delta\Phi(d)$ of F4TCNQ films. The Cu and diamond data are from Fig. 2 of ref. 22 and Fig. 3.5b of ref. 3, respectively. The curve is Eq. 10 with the $d_{ML} = 4$ Å, $\Phi_0 = 6.3$ eV and the indicated $\rho(0)$ and $\rho(1)$.

Photoemissions $E_1$ and $E_2$ indicate F4TCNQ$^-$ at sub ML coverage, while $E_3$ indicates neutral molecules. The three features sometimes coexist,[9,24] with $E_3$ gaining and $E_1$, $E_2$ losing intensity with increasing film thickness. The CT model assumes weakly perturbed A or A$^-$ in the trial function $|\psi(\theta)\rangle$, at least for the π-MOs that are mixed by CI. The electronic spectrum of the model at any θ is a superposition of an ionic fraction $\rho(\theta)$ and a neutral fraction $1 - \rho(\theta)$. The CT model implies the coexistence of A$^-$ and A at sub ML coverage, a point whose confirmation is a matter of resolution.

## 4. Discussion

The CT fit in Fig. 3 of vacuum level shifts is preliminary for several reasons. First, more accurate $\Delta\Phi(d)$ data are assured if profiles become the focus rather than the existence and magnitude of $\Delta$. Second, there may be better interface inputs such as $d_{ML}$ or $A_0$. Third, the CT model can be generalized if warranted by good fits. The principal conclusion is that initial adsorption has considerable ionic character with $\rho(0) \sim 0.5$ that decreases to $\rho(1) \sim 0.1$ in a full ML. The precise value of $\rho(\theta)$ may change on considering other contributions. Since $\Delta\Phi(d)$ is linear in coverage for noninteracting dipoles, deviations from linearity indicate $V > 0$ in the CT model or a depolarization field in the Topping model. Profiles with large deviation from linearity bring out differences.

Most m-O interfaces have $\Delta < 0$ and electron transfer from molecule to metal. The vacuum level shifts[26] of $\alpha$-NPD on Au are shown in Fig. 4 up to $d = 20$ Å. D = $\alpha$-NPD (N,N'diphenyl-N,N'bis(1-naphthyl)-1,1'biphenyl-4,4'diamine) is not planar and its orientation on the surface is not known. Its molecular volume is almost twice that of F4TCNQ, but the estimated $d_{ML} = 11$ Å gives a similar $A_0$ and larger $\Phi_0 = -12.0$ eV. A flat $-\Delta\Phi(d)$ profile following a steep initial increase is typical of other m-O interfaces[12,13,27] and points to strong repulsion. The CT model fits the profile with $\rho(0) = 0.60$ and $\rho(1) = 0.071$ for $\mu_0 = eb$ and $b = d_{ML}/2 + 1.5$ Å for the atomic radius of Au. Larger b and smaller $A_0$ increases dipole-dipole repulsion in Fig. 1. CT in the ML is about nine times less than at $\theta \sim 0$.

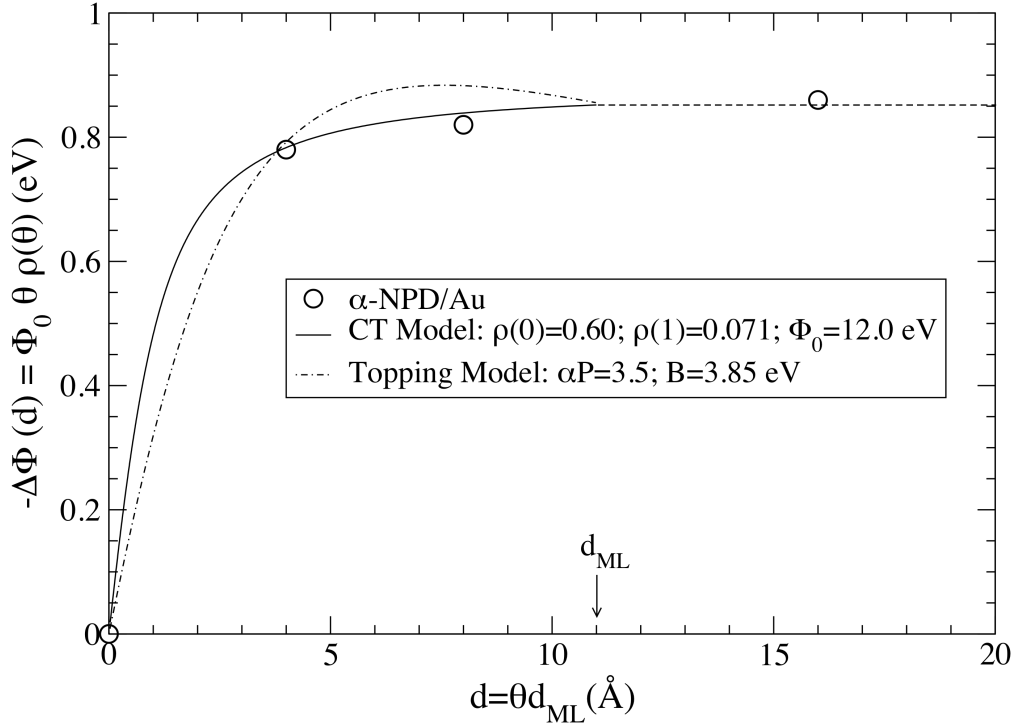

Fig. 4. Work function shift $-\Delta\Phi(d)$ of α-NPD films from ref. 26. The CT fit is Eq. 10 with the indicated $\rho(0)$, $\rho(1)$ and $\Phi_0$. The Topping fit is Eq. 11 with $\alpha P = 3.5$ and B = 3.9 eV.

The Topping model works well for adatoms on semiconductors.[19] Its parameters are the dipole $p_0$ and polarizability α of an adatom on the surface. Topping[20] first scaled the surface structure in Fig. 1 as $r(\theta) = r_0\theta^{-1/2}$. The depolarization field $F(\theta)$ is found self-consistently for induced dipoles $\alpha F(\theta)$. The net dipole $p(\theta)$ and hence $\Delta\Phi_T(\theta)$ is[19]

$$\Delta\Phi_T(\theta) = -B\theta/(1+\alpha P\theta^{3/2}) \qquad (11)$$

where $P$ contains the sum over $r^{-3}$ in Eq. 6 and B = $-4\pi e p_0/A_0$. The dashed line in Fig. 4 has $\alpha P = 3.5$ and B = 3.85 eV in Eq. 11. It does not produce an extended flat region. The shift clearly goes as $\theta^{-1/2}$ when $\alpha P$ is large, and $\alpha P < 2$ is required to avoid a maximum at $\theta < 1$. Although $\Delta\Phi_T(\theta)$ with $\alpha P < 2$ differs from the CT profile, small differences from linearity call for accurate profiles. The CT model's $\Gamma(\theta)$ in Eq. 7 has a similar $V\theta^{3/2}$ factor, but its increase is limited by decreasing $\rho(\theta)$. Rearrangement of Eq. 7 with $\Gamma(0) =$ 0 immediately shows that $\theta\rho(\theta)$ is constant for large V up to small corrections in $\rho(\theta) \ll 1$. Hence $\Delta\Phi(d)$ profiles can in principle discriminate among models.

As noted in the Introduction, $\Delta\Phi(\theta)$ profiles indicate charge redistribution but not the origin of surface dipoles. The present model invokes CT, the Topping model invokes polarizability, the pillow effect[28,13] invokes the compression of the metal's electron density on adsorption. Since the pillow contribution is always $\Delta < 0$, the surface dipole of F4TCNQ films contains a larger $\Delta > 0$ contribution. Also, local compression implies a linear profile while the observed profiles in Fig. 3 or 4 require interactions between adsorbed molecules. The $\Delta(d)$ profiles of all published m-O interfaces[12,13,27] are nonlinear to some extent. The pillow effect for $p_0$ plus a depolarization field amounts to a Topping model with $p_0$ due to compression.

In principle, the polarizability tensor $\alpha$ of an adsorbed molecule can be added to variable dipoles $\mu_0\rho(\theta)$ in the CT model. The CT dimer has polarizability $\alpha_d$ normal to the surface. An electric field F along $\mu_0$ changes $E(A^-S^+)$ by $-ebF$ and leads to $\rho(F)$. Charge redistribution within a dimer gives[29]

$$\alpha_d = eb\left(\frac{\partial\rho}{\partial F}\right)_0 = \frac{e^2b^2}{2(\Gamma^2+2)^{3/2}} \qquad (12)$$

with $\Gamma(\theta)$ defined in Eq. 7. The $\alpha_d$ contribution depends on $\theta$ and is largest at $\Gamma(\theta_1) = 0$ where CT is most sensitive to F. The remainder $\alpha' = \alpha - \alpha_d$ is due to charge redistribution that is beyond a CT dimer. The same logic applies to organic molecular crystals with anisotropic $\alpha$ tensors. Fractional atomic charges[30] yield $\alpha_C$, which is part of $\alpha = \alpha_C + \tilde{\alpha}$, and the remainder $\tilde{\alpha}$ is associated with atomic polarizability; $\alpha_C$ is typically large and may even exceed[31] $\alpha$, in which case $\tilde{\alpha} < 0$. In contrast to independent molecular computation of $\alpha$, however, the polarizability of an adsorbed molecule is a model parameter. We may or may not choose CT to represent charge redistribution. Once the choice is made, however, the depolarization factor in Eq. 11 becomes $\alpha' = \alpha - \alpha_d$ instead of $\alpha$ and the shift $\Delta\Phi(\theta)$ in Eq. 10 is multiplied by $1/(1 + \alpha'P\theta^{3/2})$.

The CT model is phenomenological. It gives a two-parameter expression for the surface dipole $\mu(\theta) = \mu_0\theta\rho(\theta)$ that is nonlinear in the coverage $\theta \leq 1$. The parameters $\Gamma(0)$ and V in Eq. 7 lead to $\rho(0)$ and $\rho(1)$ for the initial and final CT. Direct evaluation $\Gamma(0)$ or t requires explicit consideration of both molecules and surfaces. Related models of ion-radical or CT salts[1,14,15] or of molecular aggregates[16,17] are based on small t. The present model is designed for D or A that are ionized on surfaces at low coverage. Large dipoles $\mu_0$ are generated in systems with quite different bonding. Some approximations, less important ones, are readily relaxed. Small $\mu_N$ and depolarization can be included at the cost of extra parameters. Charges at finite b can be used instead of point dipoles. Films thicker than one ML can be considered. The $\theta^{3/2}$ dependence of dipole-dipole repulsion is required, but not the magnitude of V or $\Gamma(0)$ or t.

Detailed analysis of m-O interfaces is possible using DFT or quantum chemistry. Theoretical interest in surface dipoles has focused on the magnitude[13] of $\Delta$, the pillow effect,[28] induced density of interface states[32] and cluster calculations[33]. Some studies have considered both[23,34] $\theta = 1/2$ and $\theta = 1$ or variable[35] $\theta$ in sub MLs with constrained structures. Direct modeling is limited to one interface at a time and starts with an optimized ML structure that is either retained for $\theta < 1$ or optimized again at $\theta = 1/2$. Direct modeling is more appropriate, indeed mandatory, for strong adsorption with the formation of covalent bonds. The results sometimes[34,35,36] map into $\Delta\Phi_T(\theta)$ in Eq. 11 with *derived* rather than model parameters. CT dimers and related solid-state models focus on small t and on collective properties as discussed above. Some shared challenges are noted below.

Rangger *et al.*[23] studied F4TCNQ bonding on Au(1,1,1) and Ag(1,1,1) at $\theta = 1$ and $\sim 1/2$ and on Cu(1,1,1) at $\theta \sim 1/2$ using DFT and supercells. They report backbonding at CN groups that accounts for significant deviations from planar F4TCNQ and hence generates a dipole $\mu_N$ (their "molecular contribution", $\Delta E_{vac}$) normal to the surface. The $\mu_N$ contribution is opposite to their CT contribution and reduces it by almost a factor of two. Such DFT analysis of σ-bonds is complementary to the CT model with $\mu_N = 0$ and weakly coupled π-electronic states. But correlated models with small t are used precisely where DFT or MO methods have difficulties. The DFT transfer[23] of 1.8 π-electrons leads to F4TCNQ$^{1.8-}$ that in our opinion is not consistent with either $E_1$, $E_2$ evidence for F4TCNQ$^-$ anions or for extensive solid-state spectra that point to radical ions rather than dianions, even in stacks with σ-bonded TCNQ$^-$ that are not planar.[37] DFT with on-site repulsion U should limit transfer to one π-electron without major changes in σ-bonding.

In a computational study, Natan *et al.*[34] analyzed reconstructed Si surfaces at coverage θ = 1/2 and 1 by *p*-substituted phenyls with the formation of Si-C σ-bonds. They find a linear dependence of the surface dipole on the gas-phase dipole $p_0$. They emphasize the collective nature of dipole formation and find good agreement with Eq. 11 with αP ~ 0.9 for the calculated α. They conclude that CT or other contributions to surface dipoles are small in these systems. Without disagreeing with the conclusion, we note such small deviations from linear Δ(d) profiles can be fit with the CT model, and that CT may be appropriate in some contexts.

Romaner *et al.*[35] studied depolarization and dielectric constants in self-assembled monolayers of 4'substituted 4-mercaptobiphenyl derivatives using DFT and related their results to Eq. 11. The ML structure is accurately known: polar molecules of length b are standing on the surface with a small tilt angle and are tightly packed with separation $r_0$ < b. The point dipole approximation of Eq. 11 requires modification,[35] as expected, for b > $r_0$ and the issue comes up whenever $r_0$ and $d_{ML}$ are comparable. They assumed that sub ML structures retain erect molecules while scaling r(θ) as $r_0 \theta^{-1/2}$. Scaling is a convenient fiction when covalent bonds are formed on adsorption, especially for standing molecules. Nevertheless, the dependence of polarization or other properties on b/$r_0$ or on θ can be extracted from such conceptual studies. Surface structure is a general concern in sub ML films.

In summary, we have modeled sub ML films of F4TCNQ in terms of CT dimers with surface states of the metal, as shown in Fig. 1. Small t mixes the neutral and ionic states and leads to variable electron transfer ρ(θ) due to dipole-dipole repulsion that is treated self-consistently. Collective CT accounts for having both A⁻ and A on the surface of F4TCNQ films and provides a general approach to profiles of work function shifts or surface dipoles. The model is phenomenological, well suited for modeling $\Delta\Phi(d)$ profiles of m-O interfaces. It is closely related to quantum cell models of donors and acceptors in π-radical solids.

**Acknowledgements**. ZGS thanks Antoine Kahn for stimulating discussions about interfaces and work function shifts. We gratefully acknowledge support for work by the National Science Foundation under the MRSEC program (DMR-0819860).

References

1. Z.G. Soos, Ann. Rev. Phys. Chem. **25**, 121 (1974); Z.G. Soos and D.J. Klein, in *Treatise on Solid State Chemistry*, Vol. III (N.B. Hannay, Ed., Plenum, New York, 1976) p. 689.
2. J.S. Miller, Ed. *Extended Linear Chain Compounds*, Vol. 3 (Plenum, New York, 1983).
3. W. Chen, D. Qi, X. Gao and A.T.S. Wee, Prog. Surf. Sci. **84**, 279 (2009).
4. W. Gao and A. Kahn, Org. Electron. **3**, 53 (2002).
5. J. Blochwitz, M. Pfeiffer, T. Fritz and K. Leo, Appl. Phys. Lett. **73**, 729 (1998); X. Zhou, M. Pfeiffer, J. Blochwitz, A. Werner, A. Nollau, T. Fritz and K. Leo, Appl. Phys. Lett. **78**, 410 (2001).
6. T. Takenobu, T. Kanbara, N. Akima, T. Takahashi, M. Shiraishi, K. Tsukagoshi, H. Kataura, Y. Aoyagi and Y. Iwasa, Adv. Mater. **17**, 2430 (2005).
7. N. Koch, S. Duhm, J.R. Rabe, A. Vollmer and R.L. Johnson, Phys. Rev. Lett. **95**, 237601 (2005).
8. W.D. Grobman, R.A. Pollak, D.E. Eastman, E.T. Maas, Jr. and B.A. Scott, Phys. Rev. Lett. **32**, 534 (1974).
9. K. Mukai and J. Yoshinobu, J. of Electr. Spectr. Rel. Phen. **174**, 55 (2009).
10. S. Masuda, H. Hayashi, Y. Harada and S. Kato, Chem. Phys. Lett. **180**, 279 (1981).

11. I. Fernandez-Torrente, S. Monteret, K.J. Franke, J. Fraxedas, N. Lorente and J.I. Pascual, Phys. Rev. Lett. **99**, 176103 (2007).
12. H. Ishii, K. Sugiyama, E. Ito and K. Seki, Adv. Mater. 11, 605 (1999).
13. J. Hwang, A. Wan and A. Kahn, Mater. Sci. Eng. R, **64**, 1 (2009).
14. Z.G. Soos, H.J. Keller, W. Moroni, and D. Nöthe, N.Y. Acad. Sci. **313**, 442 (1978).
15. Z.G. Soos and A. Painelli, Phys. Rev. B **75**, 155119 (2007).
16. A. Painelli and F. Terenziani, J. Amer. Chem. Soc. **125**, 5624 (2003); F. Terenziani and A. Painell1, Phys. Rev. B **68**, 165405 (2003).
17. A. Painelli, F. Terenziani and Z.G. Soos, Theor. Chem. Acc. **117**, 915 (2007).
18. D.M. Newns, Phys. Rev. **178**, 1123 (1969).
19. W. Mönch, *Semiconductor Surfaces and Interfaces*, Springer Series in Surface Sciences 26 (Springer-Verlag, Berlin, 1993) p. 236.
20. J. Topping, Proc. Roy. Soc. A **114**, 67 (1927).
21. B.J. Topham, M. Kumar and Z.G. Soos, to be published.
22. L. Romaner, G. Heimel, J.L. Brédas, A. Gerlach, F. Schreiber, R.L. Johnson, J. Zegenhagen, S. Duhm, N. Koch and E. Zojer, Phys. Rev. Lett. **99**, 256801 (2007).
23. G.M. Rangger, O.T. Hofmann, L. Romaner, G. Heimel, B. Bröker, R.P. Blum, R.L. Johnson, N. Koch and E. Zojer, Phys. Rev. B **79**, 165306 (2009).
24. S. Braun and W.R. Salaneck, Chem. Phys. Lett. 438, 259 (2007).
25. M.J. Frisch, *et al*. Gaussian 03, Revision C.02, (Gaussian, Inc. Wallingford, CT, 2003).
26. W. Gao and A. Kahn, J. Appl. Phys. **94**, 359 (2003).
27. H. Ishii, N. Hayashi, E Ito, Y. Washizu, K. Sugi, Y. Kimura, M. Niwano, Y. Ouchi and K. Seki, Phys. Stat. Sol. (a) 201, 1075 (2004); B. Broker, R-P Blum, L. Beverina, O.T. Hofmann, M. Sassi, R. Ruffo, G.A. Pagani, G. Heimel, A. Vollmer, J. Frisch, J.P. Rabe, E. Zojer and N. Koch, Chem. Phys. Chem. 10, 2947 (2009).
28. H. Vasquez, Y.J. Dappe, J. Ortega and F. Flores, J. Chem. Phys. **126**, 144703 (2007).
29. A. Painelli and Z.G. Soos, Chem. Phys. **325**, 48 (2006).
30. E.V. Tsiper and Z.G. Soos, Phys. Rev. B **64**, 195124 (2001).
31. E.V. Tsiper and Z.G. Soos, Phys Rev. B **68**, 085301 (2003).

32. E. Abad, J. Ortega and F. Flores, J. Vac. Sci. Technol. B **27**, 2008 (2009); E. Abad, J. Ortega and F. Flores, J. Vac. Sci. Technol. B **27**, 2008 (2009).
33. I. Avilov, V. Geskin and J. Cornil, Adv. Funct. Mater. **19**, 624 (2009).
34. A. Natan, Y. Zidon, Y. Shapira and L. Kronik, Phys. Rev. B **73**, 193310 (2006).
35. L. Romaner, G. Heimel, C. Ambrosch-Draxl and E. Zojer, Adv. Funct. Mater. **18**, 3999 (2008).
36. M. Piacenza, S. D'Agostino, E. Fabiano and F. Della Sala, Phys. Rev. B 80, 153101 (2009).
37. R.H. Harms, H.J. Keller, D. Nöthe, M. Werner, D. Gundl, H. Sixl, Z.G. Soos, and R.M. Metzger, Mol. Cryst. Liq. Cryst. **65**, 179 (1981).